\documentclass[seceq]{ptptex}

\usepackage{graphicx}
\makeatletter
\@ifundefined{lesssim}{\def\lesssim{\mathrel{\mathpalette\vereq<}}}{}
\@ifundefined{gtrsim}{\def\gtrsim{\mathrel{\mathpalette\vereq>}}}{}
\def\vereq#1#2{\lower3pt\vbox{\baselineskip1.5pt \lineskip1.5pt
\ialign{$\m@th#1\hfill##\hfil$\crcr#2\crcr\sim\crcr}}}
\makeatother
\newcommand{\beq}{\begin{equation}}
\newcommand{\eeq}{\end{equation}}
\newcommand{\beqn}{\begin{eqnarray}}
\newcommand{\eeqn}{\end{eqnarray}}
\newcommand{\varep}{\varepsilon}

\title{%        %You can use \\ for explicit line-break
A Formation Mechanism of Collapsar Black Hole -- early evolution phase --%
}

\author{%       %Use \scshape  for the family name
Yuichiro \textsc{Sekiguchi} %
 and Masaru \textsc{Shibata}
}

\inst{%     %Affiliation, neglected when [addenda] or [errata]
Graduate School of Arts and Sciences, University of Tokyo,
Tokyo 153-8902, Japan
}

\abst{%         %this abstract is neglected when [addenda] or [errata]
The latest studies of  massive star evolution indicate that
an initially rapidly rotating star with
sufficiently low metallicity can produce a rapidly rotating, massive
stellar core that could be a progenitor of long-soft gamma-ray bursts
(LGRBs).
Motivated by these studies, we follow the collapse of a rapidly
rotating massive stellar core to a 'collapsar' black hole (BH)
surrounded by a massive, hot accretion disk performing
fully general relativistic simulations.
We focus on the general relativistic dynamics of the collapse, and
the relevant microphysics is treated in a qualitative manner.
The simulations are performed until the system consisting of the BH and
the disk has relaxed to a quasi-stationary state. 
A novel mechanism found in this study is that strong shock waves are
formed at the inner part of the disk after the formation of the BH. 
These shock waves propagate mainly along the rotation axis, heating the
disk and sweeping materials around the rotational axis, and thereby forming
a low density region. 
The temperature of the disk is high enough for copious neutrino
emission. 
All these features indicate that the direct formation of a rapidly
rotating BH is a promising source of LGRBs even in the absence of
strong magnetic fields.
}

\begin{document}

\maketitle

\section{Introduction}
The observed association of afterglows of long gamma-ray bursts
(LGRBs) with
star forming regions in galaxies,\cite{Goro03,CHG03,Jako05} 
%(Gorosabel et al. 2003; Christensen et al. 2004; Jakobsson et al. 2005),
such as the connections 
between GRB980425 and the Type Ib/c supernova SN1998bw,
\cite{Gala98,Kul98}
%(Galana et al. (1998); Kulkarni et al. (1998)),
between GRB030329 and the Type Ic supernova SN2003dh,
\cite{Hjorth03,Stanek03,Kawabata}
%(Hjorth et al. 2003; Stanek et al. 2003; Kawabata et al. 2003)
and 
between GRB060218 and the Type Ic supernova SN2006aj,
\cite{Modja06,Soll06}
%(Modjaz et al. (2006); Sollerman et al. (2006)), 
(see also Ref.~\citen{Zeh04} for other suggested associations between
LGRBs and core-collapse supernovae),
%(e.g., Zeh et al. (2004,2005)).
suggested that at least some LGRBs are associated
with supernovae and the core collapse of massive stars.
Supported by these evidences, the collapsar scenario, 
in which the central engine of LGRBs is composed of a rotating 
black hole (BH) surrounded by a massive, hot accretion disk,
\cite{Woosley93,Paczyn98,MacW99}
%(Woosley 1993; Paczy\'nski 1998; MacFadyen \& Woosley 1999),
is currently the most favored model for LGRBs.
It requires the progenitors to be rotating rapidly enough that 
an accretion disk can be formed around the BH.
Also, the relativistic jets that are to produce the LGRB have to 
reach the stellar surface. This constrains the size of the progenitor; 
the progenitor should not have an extended hydrogen envelope.
\cite{MacW99,Zhang} \ 
%(MacFadyen \& Woosley 1999; Zhang \& Woosley 2004).
Thus, the progenitors of LGRBs are now believed to be 
rapidly rotating massive Wolf-Rayet (WR) stars.

However, WR stars are known to be accompanied by strong stellar
winds driven by radiation pressure which lead to a rapid spin-down of
the cores of the WR stars.\cite{Langer98} \  
Some authors\cite{Podsiadlowski04,FH05}
%(Podsiadlowski et al. 2004; Fryer \& Heger 2005) 
have proposed binary-evolution models for producing rapidly rotating
progenitors of LGRBs. 
On the other hand, 
Yoon and Langer and Woosley and Heger
have recently discovered\cite{YL05,YLN06,WH06} 
that even a single star can 
fulfill the requirements of
the collapsar scenario if it is initially rapidly rotating 
($\gtrsim 50$\% of the Keplerian velocity at the equatorial surface) and
of low metallicity ($Z/Z_{\odot} \lesssim 0.1$).
The mass loss is suppressed by the low metallicity.
\cite{VK05}\ 
%(Vink \& de Koter 2005). 
The rapid rotation results in a short
mixing timescale, which leads to a chemically homogeneous state 
throughout the hydrogen burning phase.\cite{Maeder87} \  
In this case, a single star can become a rapidly rotating WR star
without losing the hydrogen envelope through the stellar wind, 
avoiding the red giant phase that otherwise would cause a
significant decrease of the core angular momentum due to
magnetic torques.
\cite{YL05,YLN06}. 

In addition to the above single-star evolutionary scenario to the LGRB 
progenitor, recent growing empirical evidence indicates that
LGRBs may prefer a low metallicity environment.
\cite{Fynbo03,Starling05,Goro05,Stanek06} \ 
%(Fynbo et al. 2003; Gorosabel et al. 2005; Starling et al. 2005;
%Stanek et al. 2006).
At least some LGRBs are likely to
be formed from a rapidly rotating WR star which is born in a low
metallicity environment and experiences chemically homogeneous
evolution.

Motivated by these latest studies, we perform axisymmetric
simulations of rapidly rotating stellar core collapse to a BH in full
general relativity. 
We explore the outcome of the
collapse of very massive, rapidly rotating WR stars in the
context of the collapsar scenario. As a first step toward more 
realistic simulations, microphysical processes are treated only in a
qualitative manner, focusing on the general relativistic hydrodynamics
of the collapse. Throughout this paper, $c$ and $G$ denote the speed
of light and the gravitational constant, respectively.

\section{Setting}
To model the core of WR stars just before the collapse, we adopt the
polytropic equation of state (EOS) $P = K\rho^{\Gamma}$, with 
$\Gamma = 4/3$, taking account of degenerate electron and radiation
pressures. Here $K$ is a constant. Rigid rotation is adopted as 
the rotation profile for simplicity. 
The maximum angular velocity (which is equal to the Kepler 
angular velocity of the equatorial surface) is assumed, since the 
progenitor (a WR star) of LGRBs is likely to be rapidly rotating. 
The central density and temperature of a pre-collapse iron core of 
massive metal-poor stars are $\rho_{c} \gtrsim 10^{9}$ g/cm$^{3}$ 
and $T_{c} \lesssim 10^{10}$ K (according to Umeda and Nomoto). 
We set the central density of the initial
conditions to  $\rho_{c} \approx 5 \times 10^{9}$ g/cm$^{3}$.
Note that the chemical potential of the electrons at such a density is 
larger than the temperature $T \approx 10^{10}$ K, and hence
the electrons are degenerate even at such a high temperature. 

According to the latest models of stellar evolution,
\cite{YL05,YLN06,WH06}
%(Yoon \& Langer 2005; Woosley \& Heger 2006a), 
rapid rotation results in a
chemically homogeneous state and leads to a large CO 
core which could produce an iron core with mass 
$\gg 2M_{\odot}$.\cite{WH06} \ 
The CO core mass $M_{\rm CO}$ can be approximately $10$, 17,
$25M_{\odot}$ for models of initial mass 20, 30,
$40M_{\odot}$, according to the result of Ref.~\citen{YLN06}.  
Taking this fact into account, we choose models with a core mass of 
$M \approx 3.5$--$4.5M_{\odot}$. This is achieved by setting 
$K \approx 9$--$10.5 \times 10^{14}$ cm$^{3}$/s$^{2}$/g$^{1/3}$. [Note
that the mass is approximately given by $4.555(K/G)^{3/2}$ for the
$\Gamma=4/3$ polytrope (see, e.g., Ref.~\citen{ST83}).] 
In this case, a BH is formed directly
as a result of the collapse for any of the chosen EOSs (see the 
next paragraph). Because we have found qualitatively the
same results for different mass models, we present the results for
$M=4.2M_{\odot}$ in the following.
In this model, the ratio of the rotational kinetic energy 
$T_{\rm rot}$ to the gravitational potential energy $W$ is 
$T_{\rm rot}/|W| \approx 0.0088$. 
The nondimensional spin parameter $q \equiv cJ/GM^{2}$ is 
$\approx 0.98$, where $M$ and $J$ denote the mass
and angular momentum of the core, respectively.

\begin{table}[t]
\caption{The parameter set  ($\Gamma_{1}$,$\Gamma_{2}$,$\rho_{\rm
  nuc}$,$\Gamma_{\rm th}$), the maximum gravitational mass $M_{\rm
  max}$, and the time $t_{\rm AH}$ at which the first apparent
  horizon is formed.}
\begin{center}
\begin{tabular}{ccccccc}
\hline \hline
Model & $\Gamma_{1}$ & $\Gamma_{2}$ & $\rho_{\rm nuc}$(g/cm$^{3}$) &
$\Gamma_{\rm th}$ & $M_{\rm max}$ ($M_{\odot}$) & $t_{\rm AH}$ (ms)
\\ \hline
A & 1.315 & 2.6  & 2.0$\times 10^{14}$ & 1.315 & 1.990 & 230 \\
B & 1.32  & 2.5  & 1.9$\times 10^{14}$ & 1.32  & 1.999 & 276 \\
C & 1.325 & 2.45 & 2.0$\times 10^{14}$ & 1.325 & 1.992 & 358 \\ 
\hline
\end{tabular}
\end{center}
\end{table}

For the evolution, we adopt a hybrid EOS 
(see, e.g., Ref.~\citen{Zwerg97}) in which the pressure consists of the 
sum of cold and thermal parts, as $ P=P_{\rm cold}+P_{\rm th} $, where 
\beqn
P_{\rm cold}=
\left\{
\begin{array}{ll}
  K_1 \rho^{\Gamma_1}, & \rho \leq \rho_{\rm nuc}, \\
  K_2 \rho^{\Gamma_2}, & \rho \geq \rho_{\rm nuc},
\end{array}
\right.\label{P12EOS}
\eeqn
($\rho_{\rm nuc}$ denoting the nuclear density) and $P_{\rm th} =
(\Gamma_{\rm th} -1)\rho (\varepsilon - \varepsilon_{\rm cold})$.
Here, $\rho$ is the rest-mass density, $\varepsilon$ and
$\varepsilon_{\rm cold}$ are the total and cold parts of the specific
internal energy, $\Gamma_{1}$, $\Gamma_{2}$, and $\Gamma_{\rm th}$ are
constants, $K_1=5\times 10^{14}$ in cgs units, and $K_2 =
K_{1}\rho_{\rm nuc}^{\Gamma_2 - \Gamma_1}$, respectively.  By choosing
a value of $\Gamma_{1}$ slightly smaller than $4/3$, the effect of
the depletion of the degenerate electron pressure due to
photo-dissociation and electron capture is qualitatively taken
into account. The specific internal energy is set to the same value as
in the case of the $\Gamma=4/3$ polytropic EOS, i.e.
$\varepsilon = 3K\rho^{1/3}$.  

The values of $\Gamma_{1}$ are chosen as 1.315, 1.32, and 1.325 in this paper.  
For the case of ordinary supernova simulations (i.e. the collapse of a stellar
core with a mass smaller than those in the present work), 
the value of $\Gamma_{1}$ may be in the range
$1.30 \le \Gamma_{1} \le 1.33$\cite{Zwerg97}.\ 
For the collapse of cores with larger masses, it has been pointed out that the
value of $\Gamma_{1}$ tends to be larger than in the lower mass
case, due to their larger entropies (see, e.g., Ref.~\citen{Nakazato}).
For this reason, we believe that the values of $\Gamma_{1}$ adopted in this
paper may be appropriate.
The values of $\Gamma_{2}$ and $\rho_{\rm nuc}$ are chosen
so that the maximum allowed gravitational mass of the cold neutron
star is $M_{\rm max} \approx 2.0M_{\odot}$, which is approximately
equal to the highest pulsar mass ever measured,\cite{Nice05} i.e., $M
= 2.1\pm 0.2M_{\odot}$.  We set $\Gamma_{\rm th} = \Gamma_{1}$ for
simplicity. With such small values, the low efficiency of the shock
heating due to the energy loss through photo-dissociation and
neutrino emission is qualitatively taking in to account .  The adopted
values for the parameter set ($\Gamma_{1}$,$\Gamma_{2}$,$\rho_{\rm nuc}$,$\Gamma_{\rm
th}$) are listed in Table I.

The temperature $T$ is estimated from 
$\varepsilon$ by solving the equation
\beq
\varepsilon = \varepsilon_{\rm gas} + \varepsilon_{\rm rad} +
\varepsilon_{\nu} + \varepsilon_{\rm cold}\, , \label{temp}
\eeq 
where $\varepsilon_{\rm gas}$, $\varepsilon_{\rm rad }$, and
$\varepsilon_{\nu}$ denote the specific internal energy of the gas,
the radiation, and the neutrinos and are functions of $\rho$ and
$T$. We use the same forms for these quantities as in
Ref.~\citen{DiMatteo02}, setting
\beqn
&&\varep_{\rm gas}={3\rho k T \over 2m_p}{1 +3X_{\rm nuc} \over 4},\\
&&\varep_{\rm rad}={11 \over 4} a T^4,\\
&&\varep_{\nu}={21 \over 8}aT^4 \Theta(\tau-\tau_{\rm crit}),\\
&&\varep_{\rm cold}={h c \over 24 \pi^3}\biggl({3\pi^2 \rho Y_e \over 
m_p}\biggr)^{4/3},
\eeqn
where $k$ is the Boltzmann constant, $h$ Planck's constant, $m_p$ 
the nucleon mass, $X_{\rm nuc}$ the mass fraction of free nucleons, 
$a$ the radiation constant, and $Y_e$ the electron fraction. 
Further, $\Theta$ denotes the Heaviside step function; i.e., for an optical
depth $\tau$ larger than the critical optical depth 
$\tau_{\rm crit} \approx 2/3$, we assume that the neutrinos 
are opaque with respect to nucleons and electrons and take into account 
their energy densities. (The definition of $\tau$ is given in \S 3.) 

Simulations were performed using an axisymmetric, full general
relativistic code.\cite{Shibata03} \  
The so-called simple excision technique\cite{AB01} 
was implemented to follow the evolution of the BH spacetime in a stable manner.
(For recent calculations, see, e.g., Refs.~\citen{ShiTa06} and \citen{SDLSS}.)
The regridding technique\cite{ShiSha02} was adopted during
the collapse. The grid size and grid spacing in the final regridding
were $(3300, 3300)$ for $(\omega, z)$ and $\Delta x \approx M/20$, where
$\omega = \sqrt{x^{2} + y^{2}}$ is the radius in cylindrical
coordinates. 

\section{Results}
\begin{figure}[ht]
\begin{center}
  \begin{minipage}[t]{0.47\textwidth}
    \includegraphics[width=5.3cm]{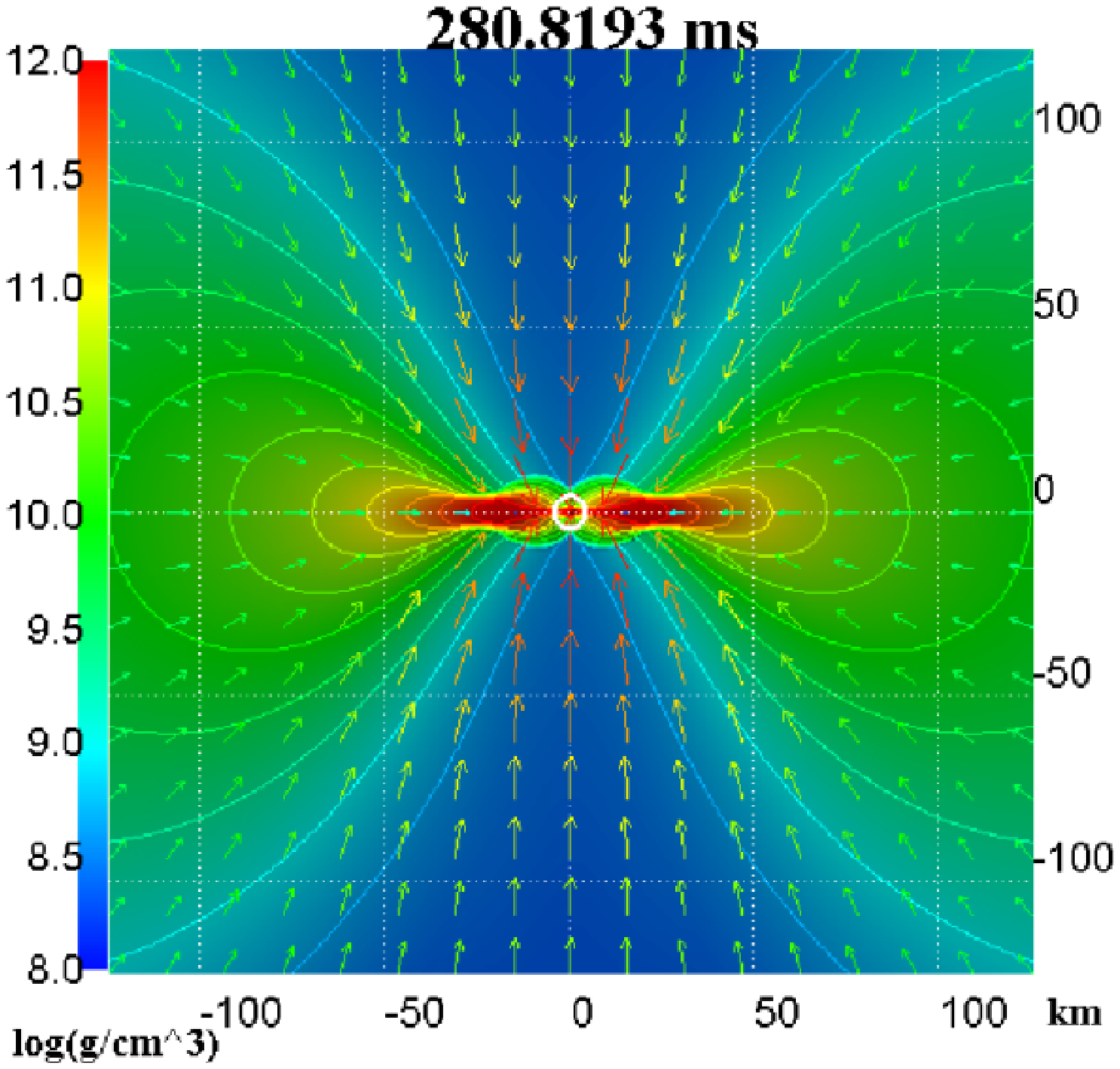}
  \end{minipage}
  \begin{minipage}[t]{0.47\textwidth}
    \includegraphics[width=5.3cm]{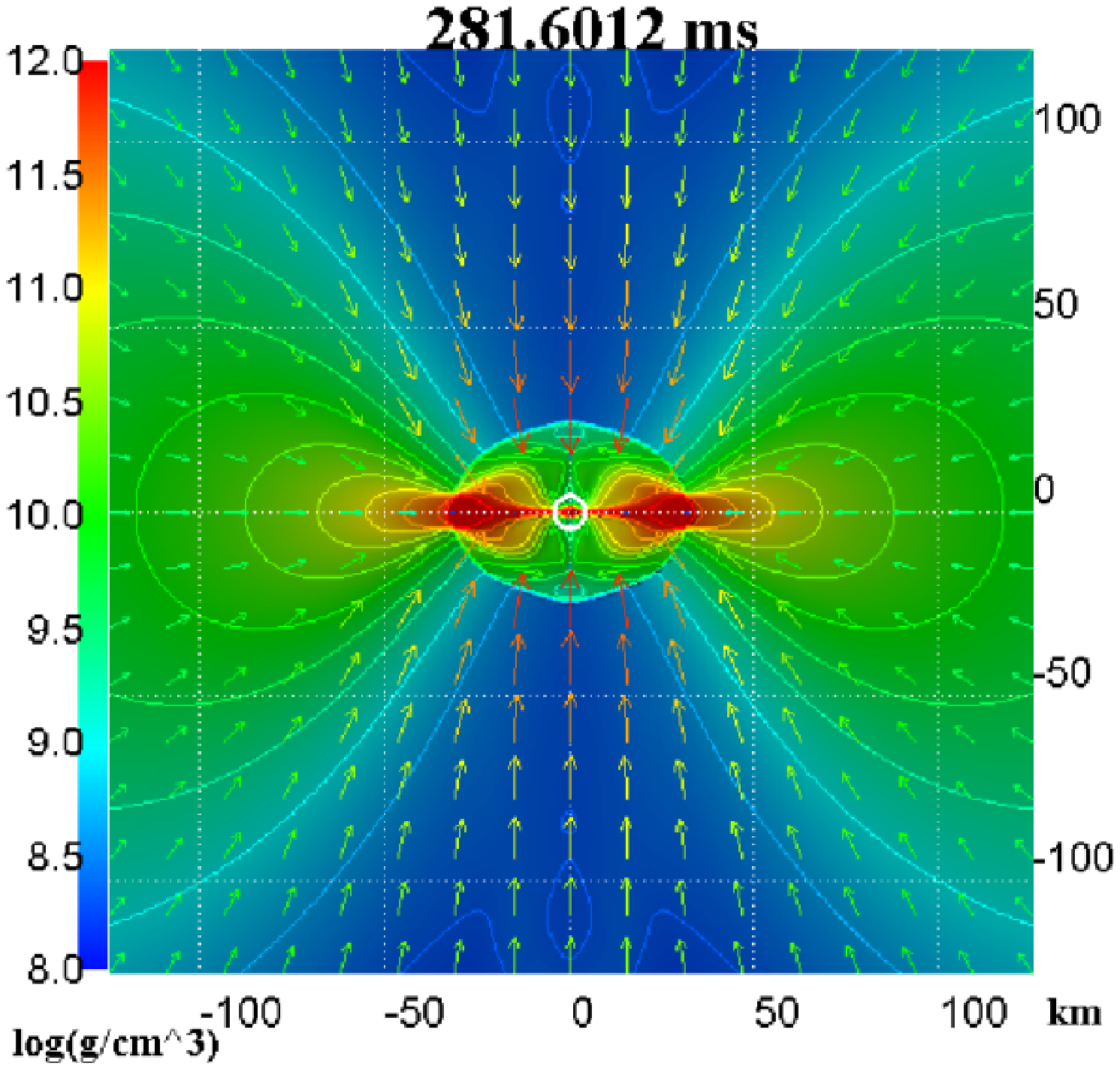}
  \end{minipage} \\
  \begin{minipage}[t]{0.47\textwidth}
    \includegraphics[width=5.3cm]{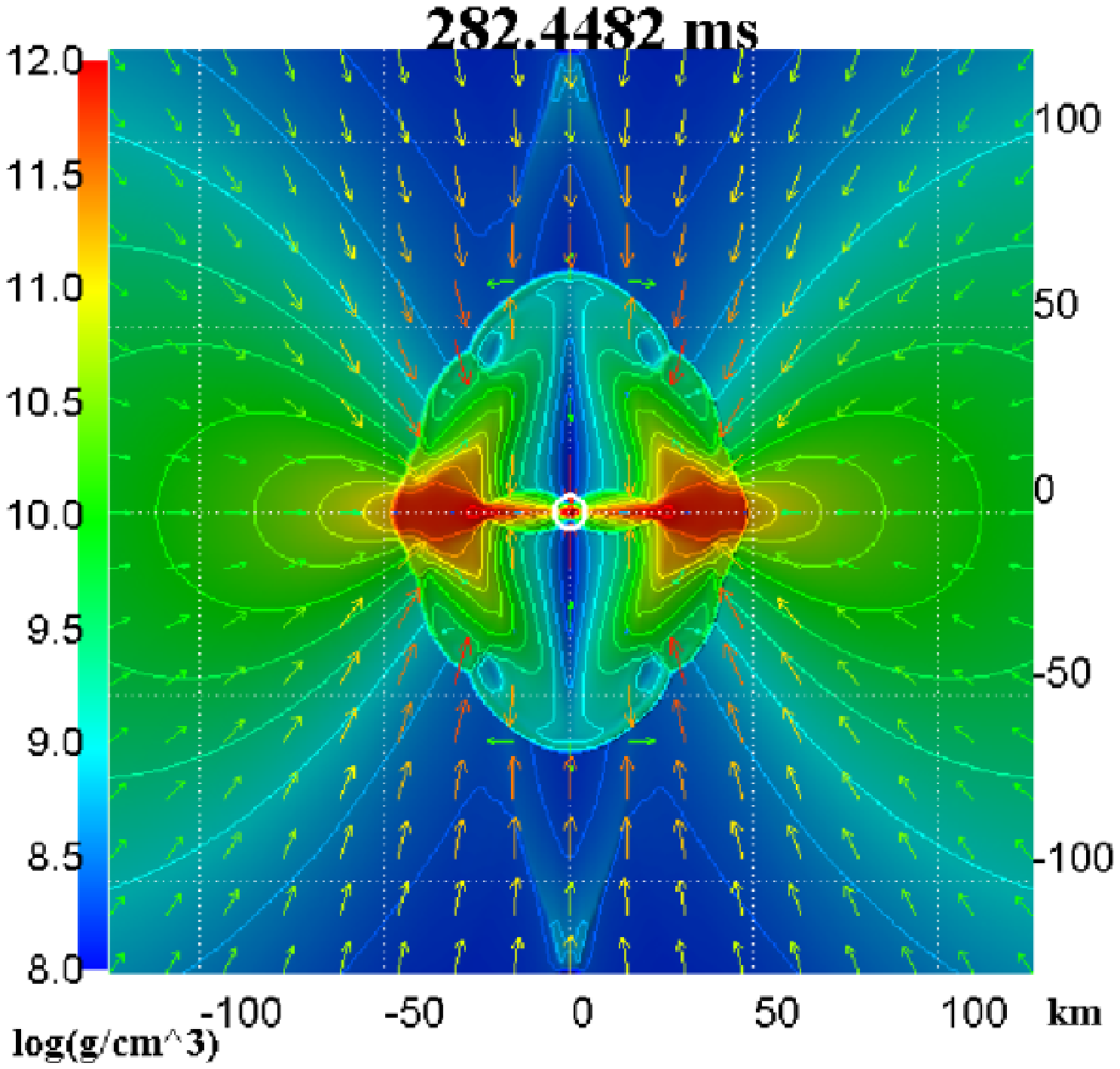}
  \end{minipage}
  \begin{minipage}[t]{0.47\textwidth}
    \includegraphics[width=5.3cm]{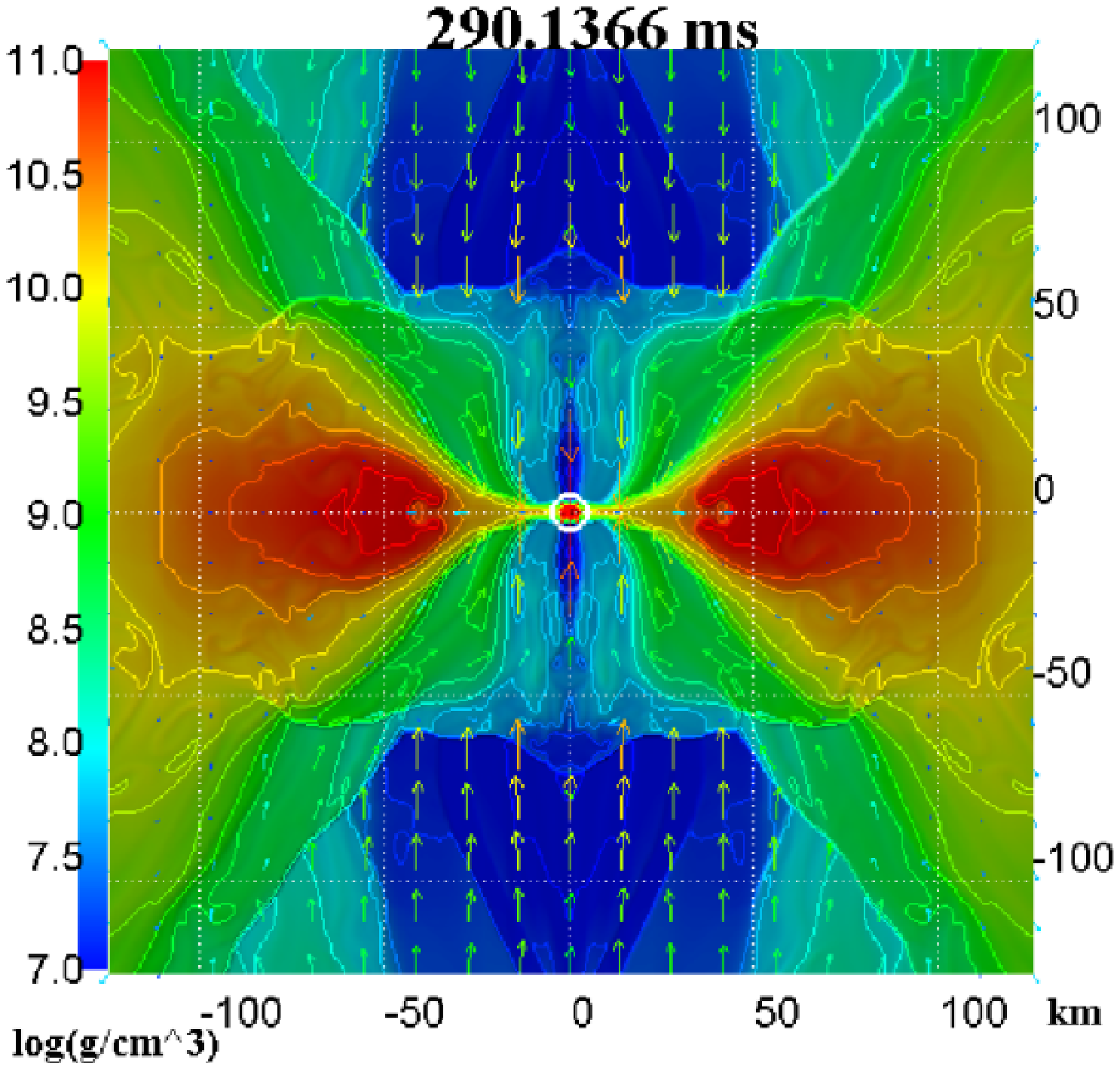}
  \end{minipage} \\
  \begin{minipage}[t]{0.47\textwidth}
    \includegraphics[width=5.3cm]{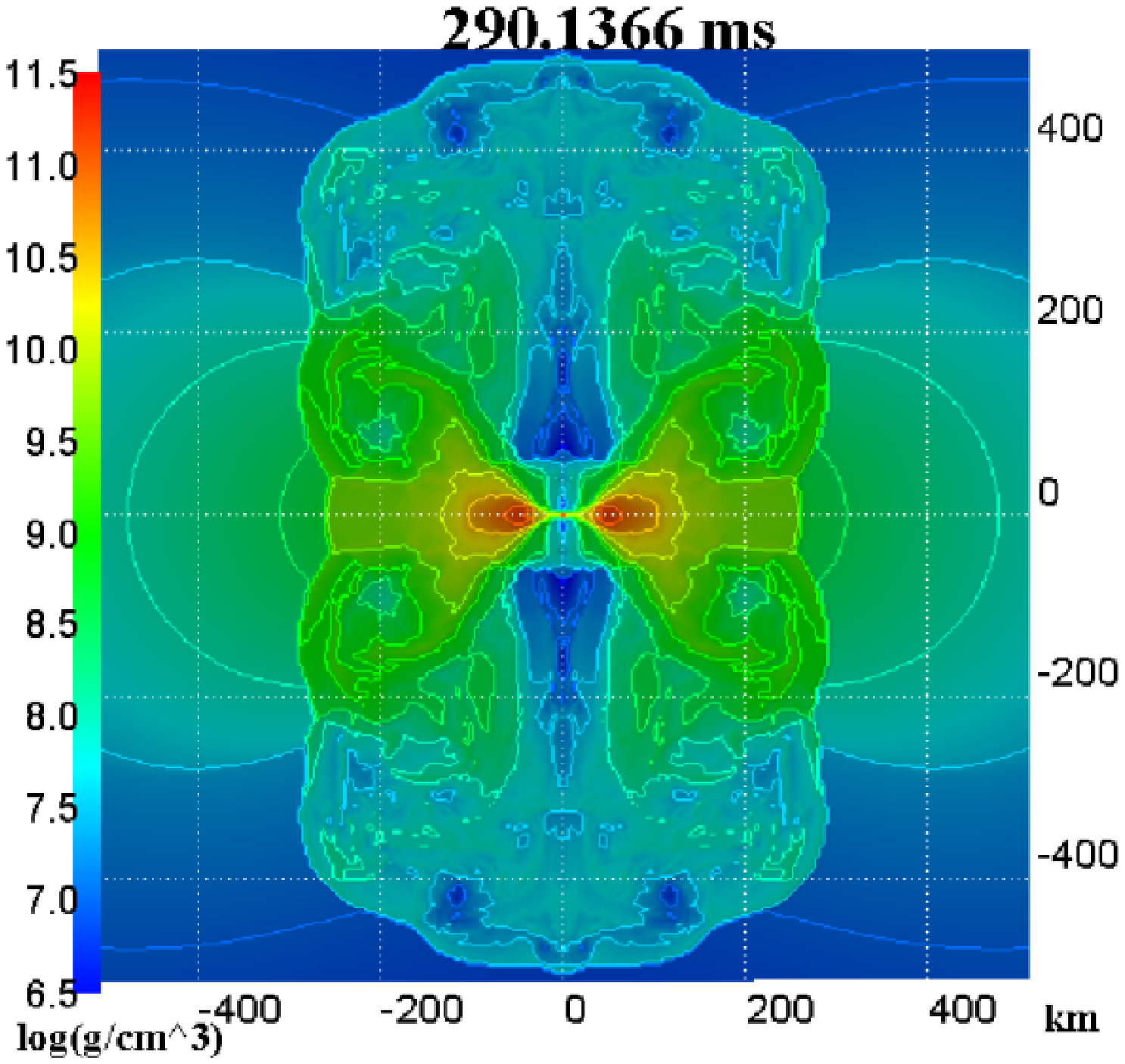}
  \end{minipage}
  \begin{minipage}[t]{0.47\textwidth}
    \includegraphics[width=5.3cm]{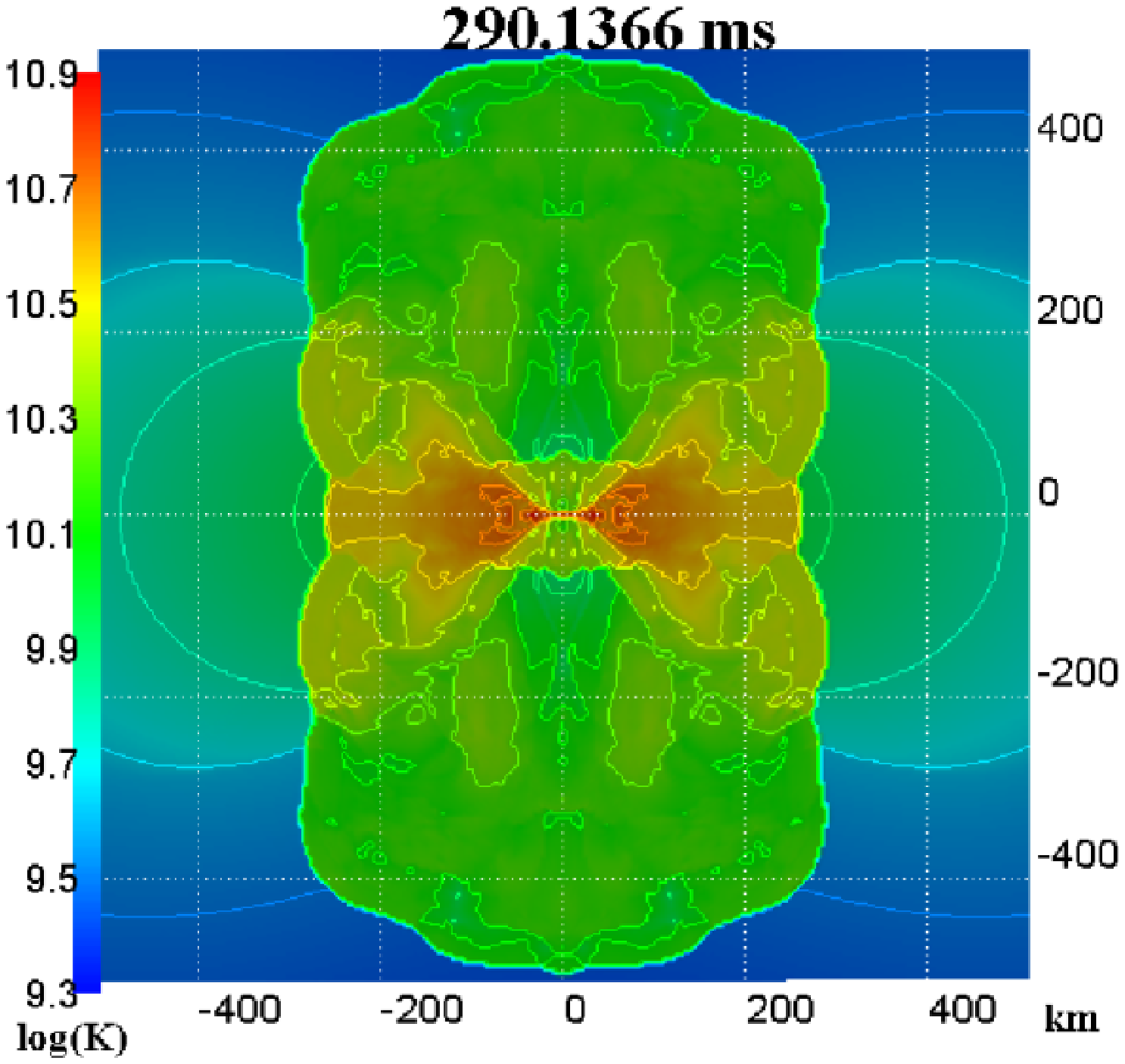}
  \end{minipage}
\end{center}
\vspace{-3mm}
\caption{Snapshots of density contours and velocity fields of the central
  region of the core in the $x$-$z$ plane at $t\approx 280.82$ms 
  (the top left panel), 281.60ms (the top right panel), 
  282.25ms (the middle left panel), and 290.14ms (the middle right panel). 
  Snapshots of density contours (the bottom left panel) and 
  temperature contours (the bottom right panel) in the $x$-$z$ plane at
  290.14 ms. All figures are for model B.
\label{fig1}}
\end{figure}
%%%%
\begin{figure}[ht]
\begin{center}
    \includegraphics[width=9cm]{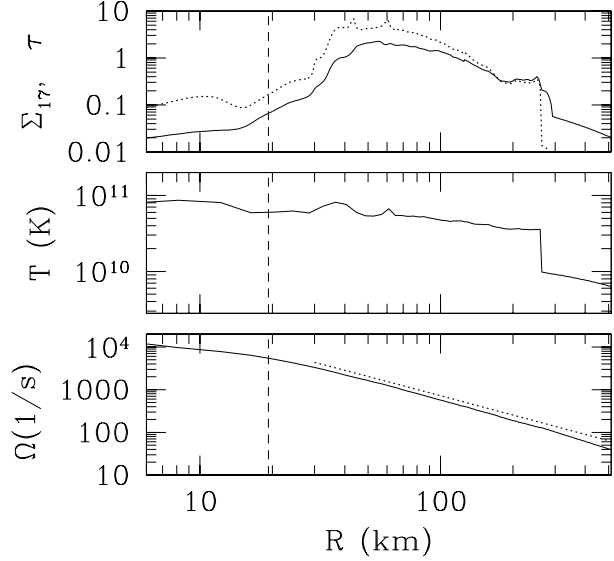}
\end{center}
\vspace{-10mm}
\caption{Various quantities characterizing the disk at 290.14 ms for model B. 
  The top panel plots the surface density 
  $\Sigma_{17} \equiv \Sigma/10^{17}$g/cm$^{2}$ (solid curve) 
  and the optical depth $\tau \approx \kappa_{\nu} \Sigma$ (dotted
  curve). 
  The middle panel plots the temperature evaluated in the equatorial
  plane. The bottom panel plots the angular velocity $\Omega$ (solid
  line) and the Keplerian angular velocity (dotted curve). 
  The vertical dashed lines at $R\approx 19$ km indicate the expected
  location of the marginally stable circular orbit around the BH.
  \label{fig2}}
\end{figure}
%\vbox{
%\vskip 0.125truecm
%\centerline{\epsfxsize=8.2truecm\epsfbox{f2.eps}}
%\vspace{-10mm}
%\figcaption[]{Various quantities of the disk at 290.14 ms for model B. 
%  The top panel shows the surface density 
%  $\Sigma_{17} \equiv \Sigma/10^{17}$g/cm$^{2}$ (solid curve) 
%  and the optical depth $\tau \approx \kappa_{\nu} \Sigma$ (dotted
%  curve). 
%  The middle panel shows the temperature evaluated in the equatorial
%  plane. The bottom panel shows the angular velocity $\Omega$ (solid
%  line) and the Keplerian angular velocity (dotted curve). 
%  The vertical dashed lines at $R\approx 19$km denote the expected
%  location of marginally stable circular orbit around the BH.
%  \label{fig2}}
%\vskip 0.125truecm  } 
%%%%%
\begin{figure}[ht]
  \begin{center}
    \includegraphics[width=9cm]{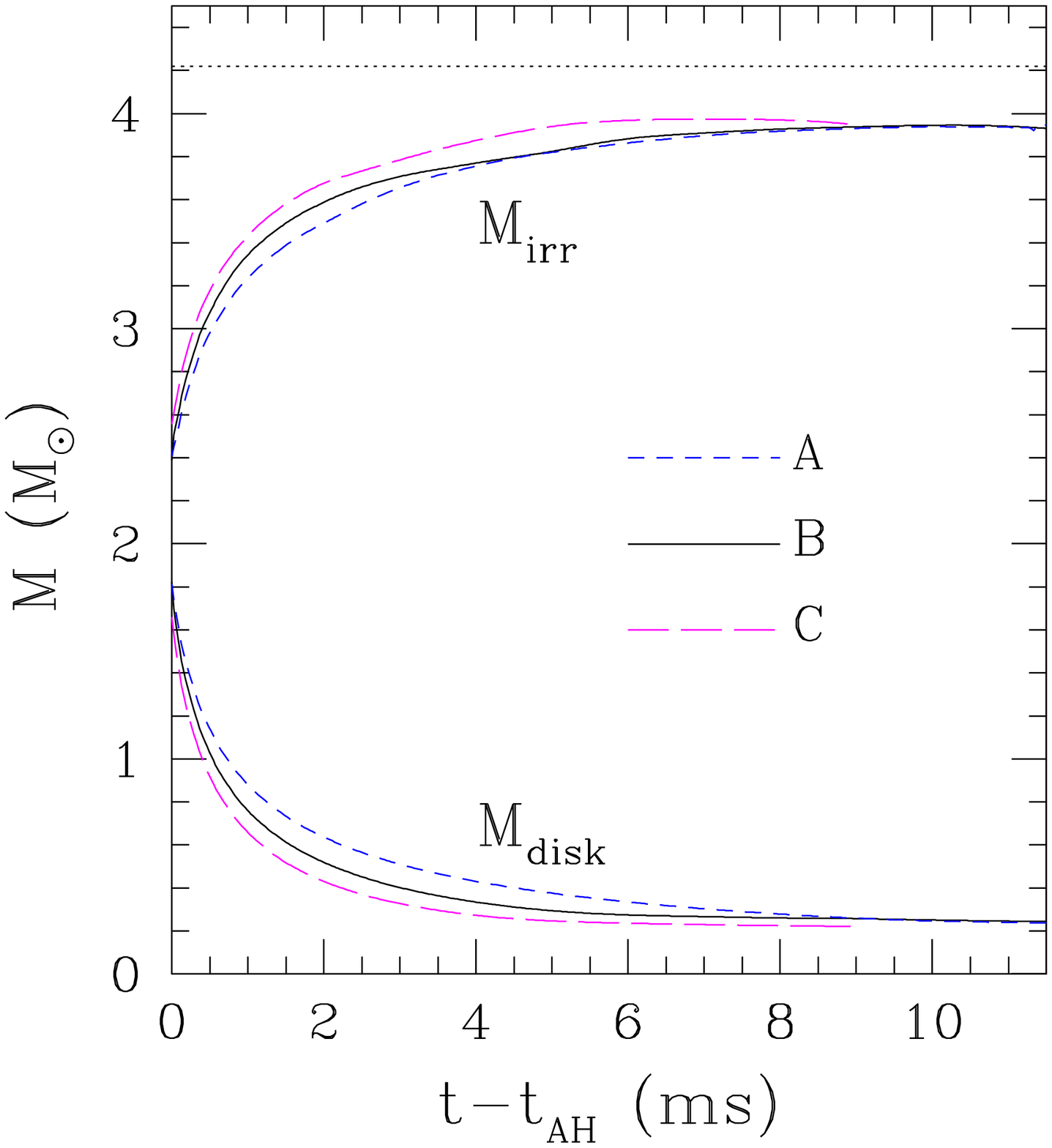}
  \end{center}
  \caption{Evolutions of the irreducible mass $M_{\rm irr}$ and the disk
    mass $M_{\rm disk}$. The long dashed, solid, and dashed lines represent
    the results for models A, B, and C, respectively. The horizontal line
    at $M\approx 4.2M_{\odot}$ indicates the total mass.\label{fig3}}
\end{figure}
%\vbox{
%\vskip 0.125truecm
%\centerline{\epsfxsize=8.truecm\epsfbox{f3.eps}}
%\figcaption[]{Evolution of irreducible mass $M_{\rm irr}$ and disk
%  mass $M_{\rm disk}$. The long dashed, solid, and dashed lines denote
%  the result for model A, B, and C, respectively. The horizontal line
%  at $M\approx 4.2M_{\odot}$ denotes the total mass.\label{fig3}} 
%\vskip 0.125truecm  } 
%

In the early stage, during which the central density of collapsing core is 
smaller than the nuclear density, the collapse proceeds in an approximately
homologous manner. 
Then, as the central density approaches the nuclear density, the collapse
around the equatorial plane is decelerated, due to the centrifugal
force, while the collapse along the rotational axis is relatively
accelerated to form a flattened structure.
Since the mass of the stellar core is much larger than the maximum allowed
mass for the given EOS, a BH is formed directly.
The formation of the BH is ascertained by finding the apparent horizon.
After the apparent horizon is formed (for the formation time, see Table I),
a centrifugally supported, thin accretion disk is formed around the BH
(see the top left panel of Fig. 1):
The disk rotates at approximately the Keplerian velocity. (see the bottom
panel of Fig. 2).
The evolution of the irreducible mass of the BH, $M_{\rm irr} \equiv
\sqrt{A/16\pi}$, and the disk mass $M_{\rm disk}$ are shown in Fig. 3.
Here $A$ is the area of the apparent horizon.

After the disk formation, shocks are formed at the inner part of the disk, 
converting the kinetic energy of the infall into  thermal energy
(See Ref.~\citen{Lee06} for discussion of a similar phenomenon.) 
The gravitational potential energy released is  
$E \sim GM_{\rm BH}M_{\rm disk}/R_{\rm ISCO} \approx 
4$--5$\times 10^{52}$ ergs, 
where $M_{\rm BH}$, $M_{\rm disk}\sim 0.1M_{\odot}$, and 
$R_{\rm ISCO} \approx 4$--$5Gc^{-2}M_{\rm BH}$ are
the black hole mass, the disk mass, and the radius of the innermost
stable circular orbit around the BH, respectively.
As the thermal energy is stored, the disk height $H$ increases.
For $H < R_{\rm ISCO}$, $H$ in the vicinity of the BH is approximately 
determined by the balance relation 
\beq
{P_{\rm disk}-P_{\rm ram} \over H} \sim 
\frac{GM_{\rm BH}\rho_{s}H}{R_{\rm ISCO}^{\ 3}}, \label{pres}
\eeq
where the left-hand and right-hand sides represent the pressure 
gradient and the gravity of the BH. The quantities  
$\rho_{\rm s}$, $P_{\rm disk}$, and $P_{\rm ram}$ are the 
disk density near the surface, the pressure inside the disk, and the
ram pressure of the infalling matter, respectively. 
Equation (\ref{pres}) provides the order of magnitude of the 
pressure as
\beq
(P_{\rm disk}-P_{\rm ram})\sim 
\frac{GM_{\rm BH}\rho_{s}H^{2}}{R_{\rm ISCO}^{\ 3}} \sim 
10^{31}\,\left(\frac{\rho_{\rm s}}{10^{11}{\rm g/cm}^{3}}\right)\,
\left(\frac{H}{R_{\rm ISCO}}\right)^{2}
\ {\rm dyn/cm}^{2}. 
\eeq
The ram pressure can be expressed as 
\beq
P_{\rm ram} \sim \rho_{\rm f}v_{\rm f}^{2} \sim 10^{30}\,
\left(\frac{\rho_{\rm f}}{10^{10}\,{\rm g/cm}^{3}}\right)
\ {\rm dyn/cm}^{2}
\eeq
where
$\rho_{\rm f}$ and $v_{\rm f}\sim (2GM_{\rm BH}/R_{\rm ISCO})^{1/2} 
\sim 0.4$--0.5$c$ are the density and velocity of the infalling matter.

The density ($\rho_{\rm disk}$) and the temperature ($T_{\rm disk}$)
inside the disk eventually increase to $\sim 10^{12}$ g/cm$^{3}$ and 
$\sim 10^{11}$ K 
(and hence, $P_{\rm disk} \sim 10^{31}$ dyn/cm$^{2}$), 
while the ram pressure ($P_{\rm ram}$) decreases to 
$\lesssim 0.1P_{\rm disk}$ ($\ll P_{\rm disk}$),
since the density of the infalling matter ($\rho_{\rm f}$) decreases to 
$\lesssim 10^{10}$ g/cm$^{3}$.
Consequently, $H$ increases to $\sim R_{\rm ISCO}$ for 
$\rho_{\rm s} \sim 10^{11}$ g/cm$^{3}$.
Then, the approximate balance relation becomes
\beq 
(P_{\rm disk}-P_{\rm ram})\sim \frac{GM_{\rm BH}\rho_{\rm s}}{H}.
\eeq
Since the binding due to the gravitational force decreases as $H$
increases, the disk expands preponderantly in the $z$-direction, 
forming strong shock waves
(see the top right and middle left panels of Fig. 1).
The propagation speed of the
shock waves is $\approx 0.5c$, i.e., mildly relativistic.
As the shock waves propagate, the mass accretion rate to the BH
significantly decreases (see Fig. 3).  
After the shock propagation, a low density funnel region is formed
around the rotational axis (see the middle right and bottom left
panels of Fig. 1).
In addition, the funnel region is surrounded by a dense, hot wall
(see the bottom right panel of Fig. 1).
The temperature in the funnel increases to
approximately the same level as that in the wall through shock
heating, and hence such a low density funnel can be supported by the
thermal pressure. 

Figure 2 displays the surface density, the optical depth,
the temperature, and the angular velocity of the disk at $t=290.14$ ms.
Due to the shock heating, the temperature increases to $T \approx $
4--9 $\times 10^{10}$ K, which is about a factor of 4--5 larger than that in
the unshocked regions (see the middle panel of Fig. 2). 
The surface density around the rotational axis
is $\Sigma \approx$ 2--6 $\times 10^{15}$g/cm$^{2}$, while it is
$\Sigma \sim 10^{17}$ g/cm$^{2}$ in the disk. This reflects the formation of the
funnel structure formation (see the top panel of Fig. 2). 
Assuming that neutrino opacity is
$\kappa_{\nu} = 7\times 10^{-17}(T/10^{11}\, {\rm K})^{2}$ 
(see, e.g., Di Matteo et al. 2002), we compute the optical depth $\tau =
\kappa_{\nu} \Sigma$. Then we find that the disk is optically
thick for a radius of  $R \approx 20$--$130$ km (see the top panel
of Fig. 2). 
The subsequent evolution of the hot disk will be determined by the 
duration of the infall and the viscous timescale, as previously 
discussed (see, e.g., Ref.~\citen{Lee05}).

Successful LGRB formation requires at least the following three
ingredients: 1. formation of an energetic fireball which yields
highly relativistic outflows; 
2. collimation of the relativistic outflows;  
3. successful penetration of the outflows through a surrounding
stellar mantle.
As discussed in the following, all these conditions are likely to be
satisfied in our model, even in the
absence of a strong magnetic field, which is often assumed (see, e.g.,
Refs.~\citen{Proga03} and \citen{Taki04}).
%Proga et al. 2003; Takiwaki et al. 2004).

In our models, a highly relativistic fireball will be produced by the
neutrino pair ($\nu \bar{\nu}$) annihilation 
around the rotational axis.
Here, let us approximately estimate the energy deposition rate
$\dot{E}_{\nu \bar{\nu}}$ through $\nu \bar{\nu}$ annihilation.
In the diffusion limit, the neutrino flux is given by 
\beq
F_{\nu} \approx \frac{7N_{\nu}}{3}
                \frac{\sigma T^{4}}{\kappa_{\nu}\Sigma},
\eeq
where $N_{\nu}=3$ is the number of neutrino species and $\sigma$ is
the Stefan-Boltzmann constant. 
The neutrino luminosity from an optically thick disk is then 
\beq
L_{\nu} \approx 2\pi R_{\rm disk}^{2} F_{\nu} \sim 
5\times 10^{53} \,T_{11}^{2} \,\Sigma_{17}^{-1}\,\left(\frac{R_{\rm
  disk}}{70\, {\rm km}}\right)^{2} \ {\rm ergs/s}
\eeq
 where $T_{11} \equiv T/10^{11}$ K  and 
$\Sigma_{17} \equiv \Sigma/10^{17}$ g/cm$^{2}$.
According to the results of Setiawan et al.\cite{SRJ04},
the expected energy deposition rate through the $\nu \bar{\nu}$ annihilation 
would then be $\dot{E}_{\nu \bar{\nu}} \sim 10^{52}$ ergs/s.
The low density funnel region above the BH will be a favorable place
for an efficient production of $ee^{+}$ pairs through $\nu \bar{\nu}$
annihilation.

The formation of a dense, hot wall surrounding the funnel will
play an important role in collimating relativistic outflows. 
In the absence of this wall, energetic outflows fail to be collimated,
as the outflows travel through the stellar mantle. 
The shock waves formed at the birth of the BH are likely to sweep the
matter along the 
rotational axis, reaching the stellar surface before the main
relativistic outflows, which will be produced in the accretion phase
(e.g., Ref.~\citen{Lee05}), are driven.

For larger values of $|\Gamma_{1}-4/3|$, the time at the onset of 
the shock wave propagation is delayed, 
because the rate of increase of the disk pressure as well as the rate of
decrease of the ram pressure are smaller, reflecting the less homologous
nature of the collapse.
This also results in the fact that it takes a longer time for the BH
to relax to a stationary state (see Fig. 3).
However this 'time delay' is at most $\sim 10$ ms for $\Gamma_{1} \gtrsim
1.315$, and the features of the funnel structure formation are
qualitatively the same for any of the adopted EOSs.

The thermal energy is stored in the inner part of the accretion disk
if the heating timescale is shorter than the timescale of the neutrino
cooling. The condition is 
$\alpha (GM\dot{m}/R) \gg L_{\nu}\,( \sim 10^{53}\,{\rm ergs/s})$,
where $\alpha \lesssim 1$ is the effective conversion
efficiency of the kinetic energy to the thermal energy through the shocks 
and $\dot{m}$ is the mass accretion rate. This gives $\dot{m} \gg
L_{\nu}R/\alpha GM \approx \alpha^{-1}M_{\odot}$ s$^{-1}$. 
According to our results, $\dot{m} \gg 10M_{\odot}$ s$^{-1}$ at least for
$(t-t_{\rm AH}) \lesssim 10$ ms.
Thus, the neutrino cooling does not play an important role in the
thermal-energy-store phase, unless the conversion efficiency is extremely
low, i.e., $\alpha \ll 0.1$.
To summarize, our results are universal, at least qualitatively, for any
of the EOSs and microphysical processes.

We also performed simulations for $0.8 \lesssim q \lesssim 1.2$, as well
as for weakly differentially rotating cases and found essentially the
same results for all these parameter values. For larger values of
$q\approx 1.2$, more collimated shock waves are formed.  For smaller
values of $q \le 0.8$, a massive disk, which is essential for
shock formation, is not formed.  For sufficiently large values of
$q>1.2$, on the other hand, a BH is not promptly formed, due to the
effect of the centrifugal force.  In this case, a BH may be formed after a
sufficient amount of angular momentum is transported. 

The black hole excision technique enables us to continue the
simulation until the BH has relaxed to a quasi-stationary state. 
As Fig. 3 shows, a seed BH with mass $M_{\rm irr} \approx
2.4M_{\odot}$ is born at first for any of the EOSs. 
This value is approximately determined by the maximum allowed mass of 
the rigidly rotating configuration for the adopted EOSs. 
The figure also reveals that approximately 95\% of the total mass is 
eventually swallowed by the BH for any of the adopted EOSs. 
Since the total angular momentum is conserved, the angular momentum 
of the BH can be indirectly estimated from that of the disk, which 
is $\sim 20$\% of the total angular momentum. Thus, 
the final spin parameter of the BH is estimated as $\approx 0.8$. 

\section{Summary}
Motivated by recently developed single-star evolutionary models of LGRB
progenitors, we performed fully general relativistic simulations for
rapidly rotating stellar core collapse to a BH. 
We found that a BH is directly formed
as a result of the collapse of a sufficiently massive progenitor. Soon
after the BH formation, a disk of density $\sim 10^{12}~{\rm g/cm^3}$
is formed around the BH. The subsequent infall of matter from outside
produces shocks at the surface of the disk, and thermal energy is
generated, heating the disk. The thermal pressure eventually
reaches $\sim 10^{31}~{\rm dyn/cm^2}$, which is much larger than the
ram pressure of the infalling matter, and hence it is used to
sustain the vertical gravitational force. Because of the increasing
thermal energy, the disk and shock front gradually expand, and when
the scale height is comparable to the disk radius, the shock front
starts to expand with mildly relativistic speed. The strong shock
waves sweep matter around the rotational axis
and heat up the disk matter to $\sim 10^{11}$ K. 
Then, a low-density ($\rho \lesssim 10^7~{\rm g/cm^3}$)
funnel region surrounded by a hot, dense wall is formed. 
Formation of such a structure in advance of the subsequent 
quasi-stationary accretion from the hot disk and the resulting
relativistic outflows will be quite favorable
for producing the fireball and LGRBs. 

In the present simulation, we did not incorporate magnetic fields.
It should be noted that even in the absence of magnetic stress,
funnel structure, which will be essential for producing collimated
jets, is formed. However, magnetic fields can further promote the 
formation of such a funnel if the field strength is larger 
than $\sim 10^{15}$ G
(i.e., the magnetic pressure is larger than $\sim 10^{29}~{\rm
dyn/cm^2}$, which is comparable to the ram pressure of the infalling
matter). Study of magnetic field effects is left as a future project.

Although our treatments of the EOS and microphysical processes in 
the present work
are approximate, we believe that the qualitative features of the BH formation and
subsequent shock formation found here will be universal. 
To confirm this process more rigorously, more detailed quantitative
studies with a realistic EOS, including relevant microphysics, is needed.
Such a study is in progress, and the results will be reported in the near 
future.

\section*{Acknowledgements}
We thank R. Takahashi and K. Maeda for valuable comments and
discussions. 
Numerical computations were performed on the FACOM VPP5000 at the data
analysis center of NAOJ and on the NEC SX-8 at YITP. This work is
supported by JSPS Research Grant for Young Scientists (No.~1611308) and
by Monbukagakusho Grant (No.~17030004).

%\appendix
%\section{First Appendix} %Empty argument \section{} yields `Appendix'. 
%
%\section{Second Appendix}

\end{document}